\documentclass[twocolumn,preprintnumbers,amsmath,amssymb]{revtex4-2}

\usepackage{siunitx}
\usepackage[T1]{fontenc}
\usepackage[utf8]{inputenc}
\usepackage[english]{babel}
\usepackage{hyperref}
\usepackage{amssymb}
\usepackage{amsmath} 
\usepackage{graphicx}
\usepackage{booktabs}
\usepackage{caption}
\usepackage{subcaption}
\usepackage{multirow}
\usepackage{verbatim}

\begin{document}
\title{Antisolvent-Assisted Growth of Centimeter-Scale CsPbBr$_3$ Perovskite Single Crystals: A Theory-Guided Approach}
\author{I. O. Simonenko$^{1,2}$}
\author{R. G. Nazmitdinov$^{1,2}$}
\author{T. N. Vershinina$^{1,2}$}

\affiliation{$^1$Joint Institute for Nuclear Research, 141980 Dubna, Moscow region, Russia.}
\affiliation{$^2$Dubna State University, 141982 Dubna, Moscow region, Russia.}

\begin{abstract}
The fabrication of large, high-quality single crystals (SCs) of all-inorganic cesium lead bromide (CsPbBr$_3$) via accessible methods remains a significant challenge. This work presents a systematic approach to optimize the antisolvent vapor-assisted crystallization (AVC) method, where the experimental design is guided by theoretical methods at each step. A synergistic 9:1 (v/v) DMSO/DMF binary solvent is selected to balance solubility and kinetics, a choice rationalized by an analysis of Gutmann's donor numbers. Subsequently, ethanol is selected as a promising antisolvent by evaluating its properties against key criteria of miscibility and diffusion rate using Hansen Solubility Parameters (HSP) and Fick's law expressed in terms of saturated vapor pressure. Within this rationally-defined chemical system, the "growth window" is experimentally mapped, identifying an optimal precursor concentration of 0.35 M and a preliminary titration step to induce a controlled metastable state. The optimized protocol consistently yields high-quality, orthorhombic CsPbBr$_3$ SCs up to 1 cm in size within one week at room temperature. The resulting crystals exhibit high crystallinity and thermal stability up to \SI{550}{\celsius}.
\end{abstract}
\maketitle

\begin{widetext}
\begin{center}
{\bf Keywords:} lead halide perovskites, all-inorganic perovskites, single crystals, CsPbBr$_3$, antisolvent vapor-assisted crystallization (AVC)
\end{center}
\end{widetext}

\section{INTRODUCTION}

The well-known operational instability of hybrid organic-inorganic lead halide perovskites (LHPs) with the general formula ABX$_3$ (where A = MA$^+$ (CH$_3$NH$_3^+$), FA$^+$ (CH(NH$_2$)$_2^+$); B = Pb$^{2+}$; X = Cl$^-$, Br$^-$, I$^-$) remains a significant barrier to their commercialization. Despite their outstanding optoelectronic properties, including high light absorption coefficients, long charge carrier diffusion lengths, and remarkable defect tolerance, the organic components in their structure can degrade under exposure to moisture, heat, and prolonged illumination~\citep{green2014emergence, he2018high, hua2022improved, chen2022high, fan2020solution, liu2022growth, gupta2022synthesis}.

This vulnerability has motivated researchers to explore all-inorganic perovskites, among which cesium lead bromide (CsPbBr$_3$) stands out as a particularly promising candidate~\citep{peng2021crystallization, sujith2023growth}. CsPbBr$_3$ combines excellent thermal and chemical stability with outstanding optoelectronic characteristics, such as a direct and tunable band gap ($\sim$\SI{2.3}{\electronvolt}), high carrier mobility, and significant radiation hardness. These properties make it an ideal material for a wide range of applications, from stable solar cells to high-energy radiation detectors~\citep{he2018high, hua2023anisotropic, zhang2018anisotropic, zhang2019high}.

To fully harness these intrinsic properties, the fabrication of large-area, high-quality single crystals (SCs) is of paramount importance, as they are free of grain boundaries and possess a lower defect density compared to their polycrystalline counterparts~\citep{stoumpos2013crystal}. Over the years, several methods have been developed for growing CsPbBr$_3$ SCs. The Bridgman method, a high-temperature melt-growth technique, has been successfully employed to produce large, high-quality ingots, enabling the fabrication of high-performance radiation detectors~\citep{he2018high, zhang2017growth, agrawal2022thermal, zhang2020enhancing, han2023thermal}. However, this method requires high temperatures ($>$\SI{600}{\celsius}), high vacuum, and sophisticated equipment, making it costly and energy-intensive. Furthermore, the large thermal expansion coefficient and structural phase transitions of CsPbBr$_3$ during cooling often induce thermal stress, leading to crack formation and reduced crystal quality~\citep{agrawal2022thermal, su2024growth, jarwal2024development}.

Solution-based methods, which operate at significantly lower temperatures, offer a more accessible and cost-effective alternative. Inverse temperature crystallization (ITC), which leverages the retrograde solubility of CsPbBr$_3$ precursors in certain polar aprotic solvents like dimethyl sulfoxide (DMSO, C$_2$H$_6$OS) or N,N-dimethylformamide (DMF, C$_3$H$_7$NO), has been widely explored in recent years~\citep{saidaminov2017inorganic, dirin2016solution, chen2022high, wang2020low, cheng2023growth}. Although the ITC method is relatively fast, it has notable drawbacks. A key limitation is its strong dependence on specific solvents and, more importantly, the difficulty in controlling phase purity. Due to the different solubilities of CsBr and PbBr$_2$, undesirable secondary phases, such as the cesium-rich Cs$_4$PbBr$_6$ or the lead-rich CsPb$_2$Br$_5$, often co-precipitate, degrading crystal quality and device performance~\citep{saidaminov2017inorganic, dirin2016solution}. The simple solvent evaporation (SE) method is straightforward and inexpensive but typically yields numerous small crystals and can suffer from non-uniform growth kinetics if not carefully controlled, for instance, with an oil bath or humidity regulation~\citep{chen2022high, di2022reveal, xu2020cspbbr3}.

Among solution-based techniques, antisolvent vapor-assisted crystallization (AVC) has emerged as a particularly promising strategy due to its simple implementation, low cost, and room-temperature operation~\citep{rakita2016low, zhang2017centimeter, ding2017high, saidaminov2015high, su2024growth}. Rakita et al.~\citep{rakita2016low} first demonstrated the growth of millimeter-sized CsPbBr$_3$ crystals using this method, though they noted the persistent issue of byproduct formation without careful precursor saturation. Later, Zhang et al.~\citep{zhang2017centimeter} refined the technique by using a diluted antisolvent to slow the diffusion rate and successfully grew centimeter-sized crystals, albeit over a prolonged period of 14 days. Ding et al.~\citep{ding2017high} also utilized the AVC method to obtain high-quality crystals for fast-response photodetectors. More recently, Su~\citep{su2024growth} showed that while room-temperature AVC is feasible, the resulting crystals can have significant surface defects, indicating a poorly controlled growth mechanism.

It is notable that many of these advances have remained largely empirical, often presenting successful "recipes" without a systematic framework that rationalizes the choice of components and process parameters based on fundamental principles of crystallization. Crucially, a holistic understanding of how precursor solvation, mass transport kinetics, and solvent-antisolvent miscibility collectively govern the final crystal quality is still lacking. Consequently, developing a rationally designed protocol for the reliable and scalable production of high-quality crystals via AVC remains an open challenge.

In this work, we demonstrate an approach that combines a robust theoretical rationale with systematic experimental optimization. This synergy between theory and experiment not only yields an improved, rapid protocol but also provides a clear framework for understanding the interplay of key variables in the AVC process, offering a reliable and reproducible pathway for fabricating high-performance CsPbBr$_3$ SCs for their broader use in advanced optoelectronic applications.

\section{Experimental Section}

\subsection{Precursor Solution Preparation}

\begin{figure*}[!htb]
	\captionsetup{font=footnotesize}
    \centering
    \includegraphics[width=0.8\textwidth]{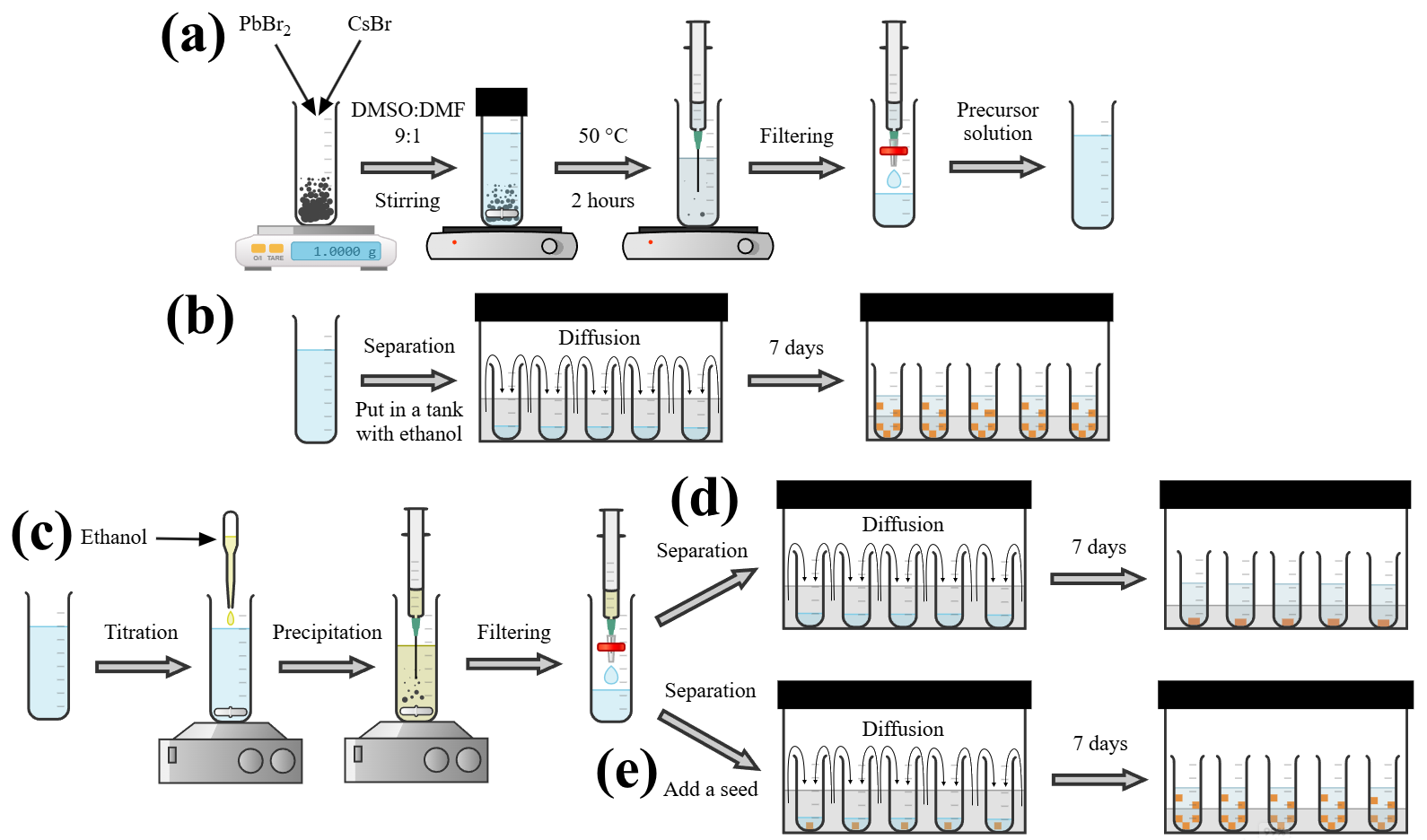}
    \caption{Schematic illustration of the experimental procedure for the synthesis of CsPbBr$_3$ SCs. (a) Preparation of the precursor solution from CsBr and PbBr$_2$ powders in a DMSO/DMF solvent mixture. (b) Crystal growth from the as-prepared solution using the AVC method (Group (a)). (c) Pre-treatment of the precursor solution via titration with ethanol and re-filtration to create a metastable state. (d) Crystal growth from the pre-treated solution using the AVC method (Group (b)). (e) Seeded crystal growth from the pre-treated solution using the AVC method (Group (c)).}
    \label{fgr:Fig_1}
\end{figure*}

Cesium bromide (CsBr, $\ge$99.8\%) and lead(II) bromide (PbBr$_2$, $\ge$99.8\%) are used as the starting materials. The precursors are mixed based on the following stoichiometric equation:
\begin{equation}
    CsBr + K\cdot PbBr_2 \rightarrow CsPbBr_3
    \label{eqn:Eq_1}
\end{equation}
where K represents the excess coefficient of PbBr$_2$. An excess of PbBr$_2$ is necessary to suppress the formation of the Cs$_4$PbBr$_6$ byproduct phase~\citep{zhang2017centimeter}. In our work, K = 1.5 is used. The mixture is dissolved in a binary solvent system of dimethyl sulfoxide (DMSO, $\ge$99.98\%) and N,N-dimethylformamide (DMF, $\ge$99.98\%) with a 9:1 volume ratio to prepare a stock solution. The solution is stirred for 2 hours at \SI{50}{\celsius} and subsequently filtered through a \SI{0.22}{\micro\meter} PTFE syringe filter to remove any undissolved particles, as depicted in Figure~\ref{fgr:Fig_1}(a).

\subsection{Synthesis of CsPbBr$_3$ Single Crystals}

Three types of precursor solutions are used for the crystal growth. The stock solution is divided into three portions. The first portion, designated as Group (a), is used as-prepared. The second and third portions are combined and subjected to a pre-treatment step involving titration with ethanol (C$_2$H$_6$O, $\ge$98\%). The titration is performed at room temperature under continuous magnetic stirring. Ethanol is added dropwise (in aliquots of $\approx$ \SI{10}{\micro\liter}) to the precursor solution. Upon each addition, a yellow precipitate forms instantaneously but redissolves rapidly upon stirring. The titration is continued until the formed yellow precipitate persists for approximately 10 seconds before dissolving. This visual endpoint corresponds to a final ethanol volume fraction of approximately 38\% (an ethanol-to-precursor volume ratio of $\approx$ 0.6:1) for 0.35 M solution. A detailed discussion of the physicochemical mechanism underlying this titration step and the associated phase diagram are presented in Section 3.3.2 and Figure~\ref{fgr:Fig_4}, accordingly. The resulting solution is then filtered through a \SI{0.22}{\micro\meter} PTFE syringe filter to remove any potential micro-nuclei and obtain a clear, metastable precursor, as shown in Figure~\ref{fgr:Fig_1}(c). This pre-treated solution is then divided into two equal volumes, designated as Group (b) and Group (c).

For the crystal growth stage, each of the three main Groups (a, b, and c) is further divided into six subgroups (1-6). This division is based on the composition of the antisolvent placed in the larger growth container, ranging from 100\% ethanol (subgroup 1) to a 50/50 (v/v) mixture of ethanol/DMSO (subgroup 6), with the DMSO content increasing in 10\% increments. This protocol results in 18 distinct experimental conditions. To ensure reproducibility, five equal aliquots from each condition are dispensed into five separate vials, which are then placed together inside the corresponding growth container.

The vials for Group (a) and Group (b) are left unchanged to study spontaneous nucleation, as illustrated in Figure~\ref{fgr:Fig_1}(b) and Figure~\ref{fgr:Fig_1}(d), respectively. For Group (c), one seed crystal ($\approx$\SI{1}{\milli\meter}$^3$) is added to each of the five vials to promote seeded growth (Figure~\ref{fgr:Fig_1}(e)). The growth of CsPbBr$_3$ SCs occurs at room temperature over one week. After the growth period, the crystals are carefully extracted from each vial, washed with DMF, and air-dried for 1 hour.

\subsection{Characterization}
The crystal structure of pulverized SCs is confirmed by X-ray diffraction (XRD) using an Empyrean (PANalytical) X-ray diffractometer with Co~K$\alpha$ radiation ($\lambda = \SI{1.7902}{\angstrom}$) and by single-crystal X-ray diffraction (SC-XRD) using a XCaliburS single-crystal X-ray diffractometer with a CCD detector and Mo~K$\alpha$ radiation ($\lambda = \SI{0.7093}{\angstrom}$). The chemical composition of SCs is studied by energy dispersive X-ray spectroscopy (EDX) with a digital scanning electronic microscope TESCAN Vega II XMU with the energy dispersive microanalysis system INCA Energy 450 (accelerating voltage 20 kV, current of the absorbed electrons on the cobalt standard 0.4 nA, with quantitative analysis based on the Cs-L$\alpha$, Pb-L$\alpha$, and Br-K$\alpha$ lines). Surface morphology and compositional homogeneity are examined using a digital scanning electronic microscope TESCAN Vega II XMU (accelerating voltage of \SI{20}{\kilo\volt}) in both Secondary Electron (SE) and Backscattered Electron (BSE) modes. UV-visible (UV-VIS) absorption spectra are recorded in the range of \SI{300}{\nano\meter} to \SI{700}{\nano\meter} on a UNICO Spectro-Quest 2804 spectrophotometer at room temperature. For optical absorption measurements, a transparent SC is selected and mechanically polished to a thickness of approximately 0.2--0.3 mm to allow for transmission measurements. Thermogravimetric analysis (TGA) is performed on a NETZSCH TG 209F1 Libra instrument to evaluate the thermal behavior of the materials. A sample weighing \SI{7.4479}{\milli\gram} is placed in an Al$_2$O$_3$ crucible and heated from \SI{30}{\celsius} to \SI{750}{\celsius} at a heating rate of \SI{10}{\celsius\per\minute}. The analysis is conducted in a dynamic argon atmosphere with a flow rate of \SI{100}{\milli\liter\per\minute}.

\section{Results and Discussion}

This section details the stepwise development and optimization of the AVC protocol for CsPbBr$_3$ SCs. Our approach is built upon a synergy of theoretical analysis and targeted experimentation, where each experimental decision, from the selection of the chemical components to the fine-tuning of process parameters, is informed by underlying physicochemical principles. We begin by (1) designing the optimal solvent system, then (2) select a suitable antisolvent based on its physicochemical properties, (3) experimentally define the "growth window" for high-quality crystallization, and (4) conclude with a comprehensive characterization of the crystals obtained via the optimized protocol.

\subsection{Rational Design of the Solvent System}

The successful growth of centimeter-scale SCs from solution is fundamentally predicated on the ability to prepare a highly concentrated and stable precursor solution. This imposes a dual, and often conflicting, set of requirements on the solvent system: it must possess sufficient solvating power to dissolve large quantities of inorganic salts (CsBr and PbBr$_2$), yet it must also provide a controllable kinetic pathway for crystallization to occur in a slow and orderly fashion.

To navigate this challenge, we employ Gutmann's Donor Number (DN) as a quantitative descriptor of a solvent's Lewis basicity and, consequently, its ability to coordinate with the Pb$^{2+}$ cation~\citep{lee2016lewis, jung2019perovskite}. A comparative analysis of two common polar aprotic solvents, DMF and DMSO, reveals their individual limitations. DMF, characterized by a moderate DN of \SI{26.6}{kcal\per\mol}, fails to provide the necessary solvating power for the inorganic precursors. Conversely, DMSO possesses a high DN of \SI{29.8}{kcal\per\mol}, enabling it to readily dissolve CsBr and PbBr$_2$ to form a high-concentration solution. However, this strong coordinating ability leads to the formation of exceptionally stable intermediate solvate complexes, [Pb(DMSO)$_x$]Br$_2$~\citep{zheng2022control}. The thermodynamic stability of these complexes translates into a high activation energy ($E_a$) for the desolvation process, which is a critical rate-limiting step for the integration of precursor ions into the crystal lattice. Consequently, a pure DMSO solution can become kinetically trapped, resisting crystallization unless a large thermodynamic driving force triggers explosive, uncontrolled nucleation, yielding polycrystalline material~\citep{li2025controlled}.

To decouple the requirements for high solubility and controlled kinetics, we devised a synergistic binary solvent strategy. The core principle is to use a solvent mixture where one component ensures high solubility while the other modulates the kinetic barrier to crystallization. Given that a high precursor concentration is non-negotiable for growing large SCs, the solvent system must be dominated by the high-DN component, DMSO. The role of the second component, DMF, is to act as a \textit{kinetic modulator}. By introducing a minor fraction of the weaker coordinating DMF, the composition of the solvate shell around the Pb$^{2+}$ ion becomes mixed. The Pb$\cdots$DMF coordination bonds are energetically weaker than the Pb$\cdots$DMSO bonds, potentially creating a lower-energy pathway for the desolvation and nucleation cascade to begin.

Guided by this analysis, we assumed that a binary solvent system dominated by high-DN DMSO for solubility, with a minor fraction of lower-DN DMF acting as a "kinetic modulator", would provide an effective balance. Consequently, a 9:1 (v/v) DMSO/DMF ratio is selected as a rational starting point for experimental optimization. This composition is intended to maintain high precursor loading, while the DMF fraction is expected to lower the kinetic barrier to nucleation. In such a way, this combination is expected to prevent the system becomes kinetically trapped~\citep{saidaminov2015high, ghosh2020antisolvents}. This theoretically-informed choice of a synergistic solvent mixture is then carried forward into all subsequent experiments.

\subsection{Antisolvent Selection}

Having established the optimal solvent system, the next critical step is the rational selection of an antisolvent. The role of the antisolvent in the AVC method is to induce a state of supersaturation by reducing the solubility of the perovskite precursors. For the growth of high-quality SCs, this process must satisfy two key requirements: it should be (1) kinetically controlled to favor growth over nucleation, and it has to be (2) spatially uniform to ensure homogeneous supersaturation. To guide our selection, we will first establish the physicochemical principles governing these requirements before systematically evaluating a range of potential candidates.

\begin{table*}
    \caption{Physicochemical properties of common antisolvents and their calculated Hansen distance ($R_a$) to the 9:1 (v/v) DMSO/DMF solvent mixture.}
    \footnotesize
    \setlength{\tabcolsep}{4pt}
    \label{tbl:Table_1}
    \begin{tabular}{lcccccc}
    \hline
    Antisolvent & Formula & $P^0_{\text{sat}}$ at \SI{20}{\celsius} & $\delta_d$ & $\delta_p$ & $\delta_h$ & $R_a$ \\
    & & (\si{kPa}) & (MPa$^{1/2}$) & (MPa$^{1/2}$) & (MPa$^{1/2}$) & (MPa$^{1/2}$) \\
    \hline
    Ethanol & C$_2$H$_6$O & 5.9 & 15.8 & 8.8 & 19.4 & 12.7 \\
    2-Propanol & C$_3$H$_8$O & 4.4 & 15.8 & 6.1 & 16.4 & 12.7 \\
    1-Butanol & C$_4$H$_{10}$O & 0.7 & 16.0 & 5.7 & 15.8 & 12.6 \\
    Dichloromethane & CH$_2$Cl$_2$ & 47.0 & 17.0 & 7.3 & 7.1 & 11.8 \\
    Chlorobenzene & C$_6$H$_5$Cl & 1.2 & 19.0 & 4.3 & 2.0 & 17.3 \\
    Diethyl ether & C$_4$H$_{10}$O & 58.7 & 14.5 & 2.9 & 4.6 & 17.8 \\
    Hexane & C$_6$H$_{14}$ & 16.0 & 14.9 & 0 & 0 & 21.9 \\
    \hline
    \end{tabular}
\end{table*}

\subsubsection{Criterion 1: Kinetic Control via Saturated Vapor Pressure}

The kinetic aspect is governed by the rate of mass transport of the antisolvent from its reservoir into the precursor solution. In the AVC setup, this rate is primarily limited by the diffusion of antisolvent vapor. According to Fick's first law, the molar flux of the antisolvent vapor ($J_{\text{av}}$) is directly proportional to its partial pressure gradient ($\frac{dP_{\text{av}}}{dx}$)~\citep{crapiste1988drying, kar1987onedimensional}:
\begin{equation}
    J_{\text{av}} = -\frac{D}{RT} \frac{dP_{\text{av}}}{dx}
    \label{eqn:Eq_2}
\end{equation}
where $D$ is the diffusion coefficient in the gaseous medium (e.g., air), $R$ is the universal gas constant, and $T$ is the absolute temperature. In our system, this gradient is established between the antisolvent reservoir, where the partial pressure is approximately equal to the saturated vapor pressure $P^0_{\text{sat, av}}$, and the surface of the precursor solution. Consequently, $P^0_{\text{sat, av}}$ becomes the key parameter controlling the crystallization rate. An optimal kinetic window should be established for this parameter. An excessively high $P^0_{\text{sat, av}}$ (>\SI{30}{kPa}) leads to uncontrolled, "flash" crystallization. This value defines the upper boundary of the window. Conversely, a vapor pressure that is too low (<\SI{2}{kPa}) could lead to impractically long synthesis durations, hindering the method's efficiency. This value defines the lower boundary.

\subsubsection{Criterion 2: Spatial Homogeneity via Hansen Solubility Parameters}

The second requirement, spatial homogeneity, depends on the thermodynamic miscibility between the antisolvent and the precursor's solvent matrix. To ensure uniform crystallization and prevent localized precipitation, the incoming antisolvent must mix effectively with the binary solvent. This property can be quantitatively assessed using Hansen Solubility Parameters (HSP), which deconstruct cohesive energy into three components: dispersion ($\delta_d$), polar ($\delta_p$), and hydrogen bonding ($\delta_h$)~\citep{hansen2007hansen, bautista2023solvent}. The compatibility between two liquids is inversely related to their distance ($R_a$) in the three-dimensional Hansen space, defined as:
\begin{equation}
    R_a^2 = 4(\delta_{d1}-\delta_{d2})^2 + (\delta_{p1}-\delta_{p2})^2 + (\delta_{h1}-\delta_{h2})^2
    \label{eqn:Eq_3}
\end{equation}
For a mixed solvent system, such as our 9:1 (v/v) DMSO/DMF, the effective HSP parameters ($\delta_{\text{mix}}$) are calculated as a volume-weighted average of the individual components' parameters:
\begin{equation}
    \delta_{\text{mix}} = \sum_{i=1}^{n} \phi_i \delta_i
    \label{eqn:Eq_4}
\end{equation}
where $\phi_i$ and $\delta_i$ are the volume fraction and the respective HSP component for the $i$-th solvent. A smaller $R_a$ distance between the antisolvent and the solvent mixture indicates better miscibility and a higher likelihood of achieving homogeneous supersaturation.

\subsubsection{Systematic Evaluation and Final Selection}

To make a rational selection, we systematically evaluate a number of commonly used antisolvents. The HSP values for all solvents are taken from the recent review by \citet{bautista2023solvent}, while the saturated vapor pressures at \SI{20}{\celsius} are obtained from Perry's Chemical Engineers' Handbook~\citep{perry2018handbook}. The results, including the calculated Hansen distance ($R_a$) to our 9:1 DMSO/DMF solvent mixture ($\delta_{mix,d}=18.3$, $\delta_{mix,p}=16.1$, $\delta_{mix,h}=10.3$), are summarized in Table~\ref{tbl:Table_1}.

A systematic analysis of the data in Table~\ref{tbl:Table_1} enables a clear, multi-criteria selection of the optimal antisolvent. Several candidates are immediately disqualified for failing to meet the fundamental criteria. Dichloromethane and diethyl ether, with extremely high vapor pressures ($\approx \SI{47.0}{kPa}$ and $\approx \SI{58.7}{kPa}$, respectively), are kinetically unsuitable. Similarly, candidates with poor predicted miscibility (evidenced by large Hansen distances) such as hexane ($R_a=21.9$) and chlorobenzene ($R_a=17.3$), are thermodynamically unfavorable. This screening process isolates the alcohols as the most promising candidates. However, 1-butanol, with a very low vapor pressure of \SI{0.7}{kPa}, presents a practical challenge, as it would lead to an impractically long synthesis time. This analysis leads to the selection of ethanol and 2-propanol as the prime contenders. Both exhibit excellent and identical miscibility ($R_a=12.7$). The key differentiator, however, is their kinetic profile. While 2-propanol's lower vapor pressure (\SI{4.4}{kPa}) leads to slower kinetics, ethanol's higher vapor pressure (\SI{5.9}{kPa}) strikes a superior balance. It is moderate enough to ensure controlled crystallization, yet sufficiently high to facilitate a time-efficient synthesis. The choice of ethanol is also strongly supported by practical considerations. Its lower toxicity, reduced cost, and widespread availability make it the ideal candidate for developing a safe, scalable, and effective synthesis protocol.

Therefore, by systematically evaluating candidates against the kinetic requirement for controlled mass transport and the thermodynamic requirement for homogeneous miscibility, ethanol is selected as the optimal antisolvent for subsequent experimental optimization.

\subsection{Defining the "Growth Window"}

\begin{figure*}[!htb]
	\captionsetup{font=footnotesize}
    \centering
    \includegraphics[width=0.8\textwidth]{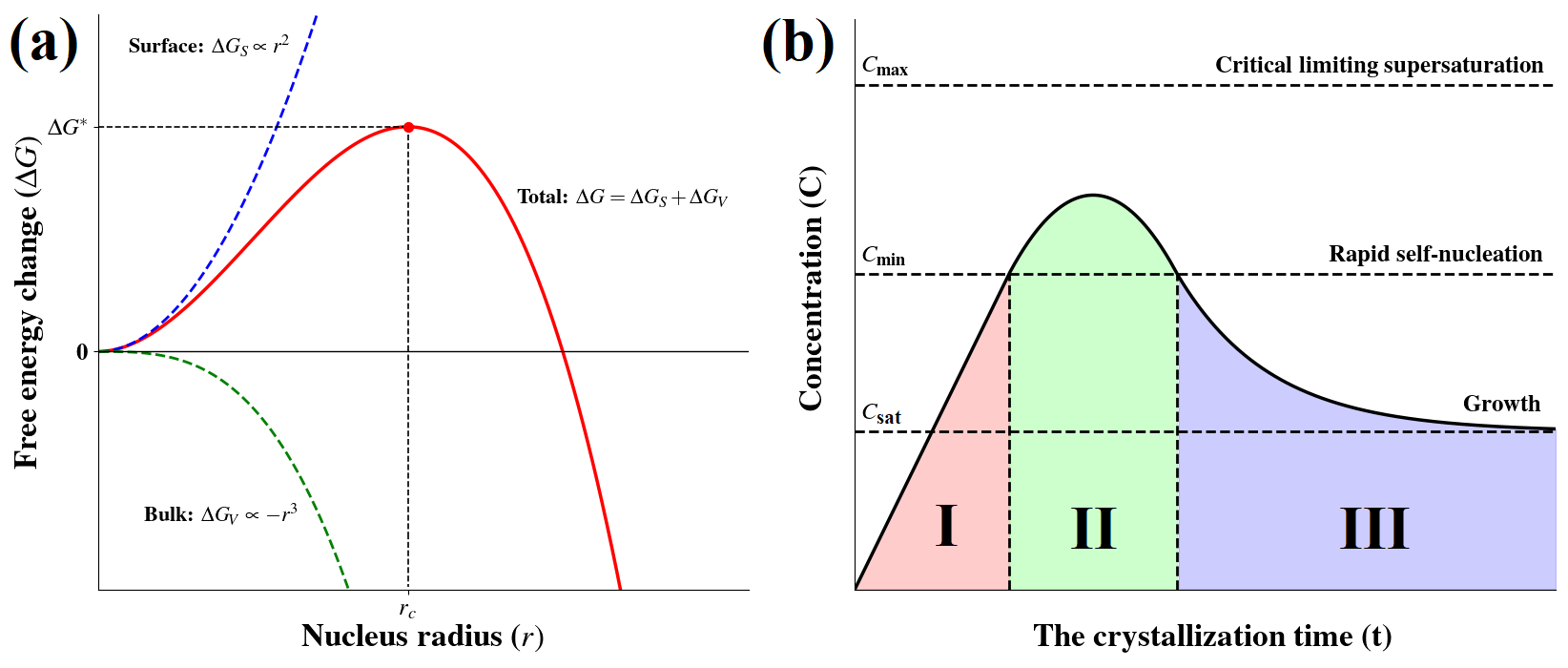}
    \caption{Theoretical framework for controlled crystallization. (a) Classical Nucleation Theory (CNT) plot showing the change in Gibbs free energy ($\Delta G$) as a function of nucleus radius ($r$). The energy barrier for nucleation ($\Delta G^*$) at the critical radius ($r_c$) must be overcome for stable growth. (b) The LaMer diagram illustrating the evolution of solute concentration ($C$) over time. Stage I is the induction period, Stage II is the burst nucleation event, and Stage III is the diffusion-limited growth of existing nuclei. Successful growth of high-quality SCs requires maintaining the system within the metastable zone between the saturation concentration ($C_{\text{sat}}$) and the minimum concentration for nucleation ($C_{\text{min}}$).}
    \label{fgr:Fig_2}
\end{figure*}

With the chemical system rationally defined (a 9:1 DMSO/DMF solvent paired with an ethanol antisolvent), the next crucial stage is to experimentally identify the optimal process parameters. While the theoretical framework guides the selection of components, it does not predict the precise conditions required for growth of high-quality SCs. The success of the synthesis now hinges on navigating the delicate interplay between thermodynamics and kinetics to operate within the so-called "growth window". This concept is visually summarized in Figure~\ref{fgr:Fig_2} and is derived from two cornerstone theories: Classical Nucleation Theory (CNT)~\citep{liao2023growth, yao2020room}, which describes the energetic barriers to forming a new phase (Figure~\ref{fgr:Fig_2}a), and the LaMer model~\citep{lamer1950theory, cao2019review}, which maps the temporal evolution of concentration during crystallization (Figure~\ref{fgr:Fig_2}b). The central idea is to operate within a narrow, metastable range of supersaturation where the conditions are favorable for the growth of existing crystals but insufficient to trigger widespread, spontaneous nucleation.

The existence of this window is a direct consequence of the fundamentally different dependencies of nucleation rate ($J$) and crystal growth rate ($V_g$) on the supersaturation ratio, $S = C/C_{\text{sat}}$, where $C$ is the current solute concentration and $C_{\text{sat}}$ is the equilibrium saturation concentration. The nucleation rate exhibits a highly non-linear, exponential dependence on supersaturation~\citep{liu2018supersaturation}, governed by the height of the energy barrier $\Delta G^*$ shown in Figure~\ref{fgr:Fig_2}a:
\begin{equation}
    J = A \exp\left(-\frac{\Delta G^*}{k_B T}\right) = A \exp\left(-\frac{16\pi\gamma^3\Omega^2}{3(k_B T)^3 (\ln S)^2}\right)
    \label{eqn:Eq_5}
\end{equation}
where $\gamma$ is the specific surface energy, $\Omega$ is the molecular volume, $k_B$ is the Boltzmann constant, and $T$ is the absolute temperature. This relationship implies a critical supersaturation threshold ($C_{\text{min}}$ in Figure~\ref{fgr:Fig_2}b), above which the nucleation rate increases explosively. In contrast, the crystal growth rate, often described by models such as the Burton-Cabrera-Frank (BCF) theory for spiral growth~\citep{shen2019defect}, follows a much slower dependence on the relative supersaturation, $\sigma = S-1$. In the low supersaturation regime ($\sigma \ll 1$), crucial for growth of high-quality SCs, the general BCF relationship simplifies to a pure quadratic dependence:
\begin{equation}
    V_g \propto \sigma^2 \tanh(\frac{B}{\sigma}) \xrightarrow{\sigma \ll 1} \sigma^2 = (S-1)^2
    \label{eqn:Eq_6}
\end{equation}
This kinetic disparity makes it possible to find a regime where $V_g$ is appreciable while $J$ is negligible. While these theories provide an essential qualitative framework, the precise quantitative boundaries of the growth window for a specific chemical system must be determined empirically. Our experimental investigation is therefore designed to systematically map out this regime by tuning three key parameters: (1) the initial precursor concentration, (2) the initial state of supersaturation, and (3) the antisolvent diffusion rate.

\subsubsection{Precursor Concentration}

\begin{figure}[!htb]
	\captionsetup{font=footnotesize}
    \centering
    \includegraphics[width=0.45\textwidth]{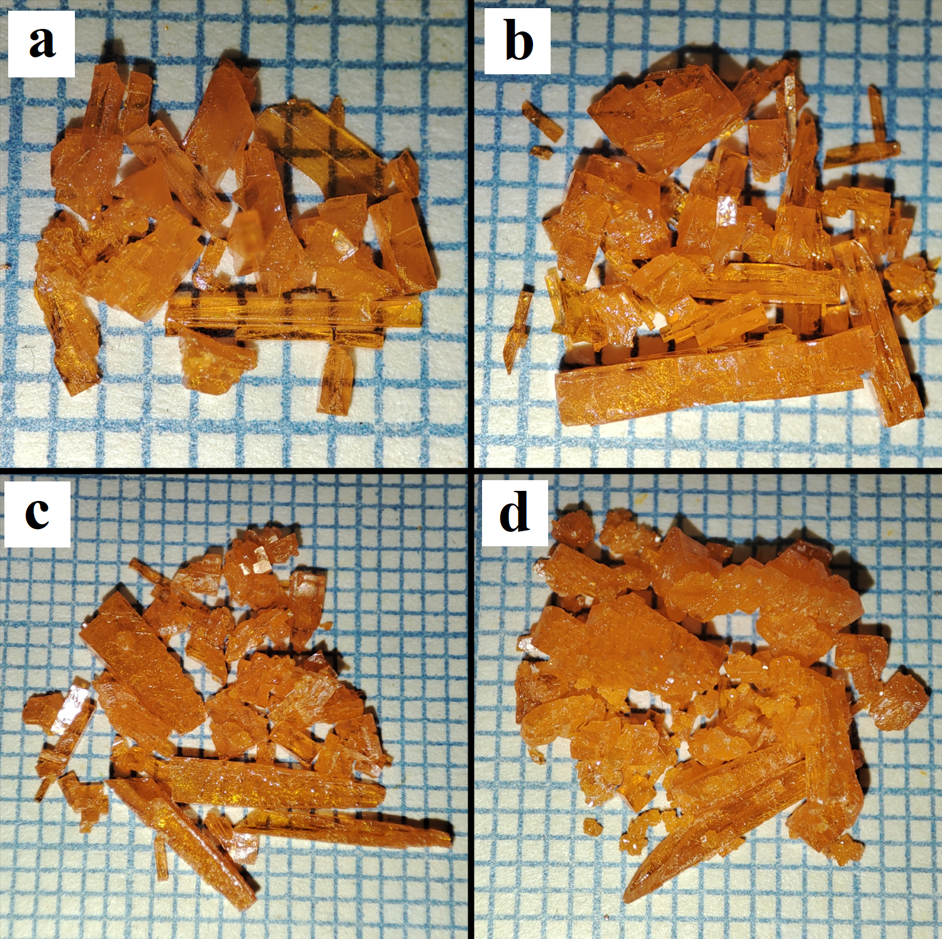}
    \caption{Photographs of CsPbBr$_3$ SCs grown from precursor solutions of varying concentrations: (a) 0.2 M, (b) 0.3 M, (c) 0.4 M, and (d) 0.5 M. The solvent used was a 9:1 DMSO/DMF mixture, and the antisolvent was pure ethanol.}
    \label{fgr:Fig_3}
\end{figure}

The initial precursor concentration is the most direct parameter for controlling the system's position relative to the supersaturation curves shown in the LaMer diagram (Figure ~\ref{fgr:Fig_2}b). According to this framework, a concentration that is too low will provide insufficient material for substantial growth, while one that is too high will cause the system to rapidly cross the nucleation threshold ($C_{\text{min}}$), leading to uncontrolled polycrystalline precipitation. To experimentally locate the optimal concentration within this theoretically-defined metastable zone, a series of experiments is conducted with concentrations ranging from 0.2 M to 0.5 M. As depicted in Figure~\ref{fgr:Fig_3}, the concentration has a profound effect on the outcome. At low concentrations (0.2 M, Figure~\ref{fgr:Fig_3}a), the total amount of solute is insufficient to produce large-volume crystals, resulting in a low yield of small SCs. As the concentration increases to 0.3 M (Figure~\ref{fgr:Fig_3}b), a noticeable improvement in crystal size is observed. However, pushing the concentration to 0.4 M and above (Figure~\ref{fgr:Fig_3}c,d) introduces a detrimental effect. A high initial concentration means that even a small influx of antisolvent vapor can cause the supersaturation ratio $S$ to rapidly exceed the critical threshold for homogeneous nucleation. This pushes the system deep into the exponential region of the nucleation rate curve (Equation~\ref{eqn:Eq_5}), triggering an avalanche of secondary nucleation events. This manifests as the formation of a large number of small, competing polycrystals and a visible degradation in the surface quality and faceting of the larger SCs. Thus, a concentration of \SI{0.35}{M} is experimentally identified as the optimal balance, providing a sufficient supply of solute for growth while maintaining a controllable level of supersaturation that avoids excessive spontaneous nucleation. This concentration is used for all subsequent experiments.

\subsubsection{Control of Initial Supersaturation and Nucleation Pathway}

\begin{figure}[!htb]
	\captionsetup{font=footnotesize}
    \centering
    \includegraphics[width=0.45\textwidth]{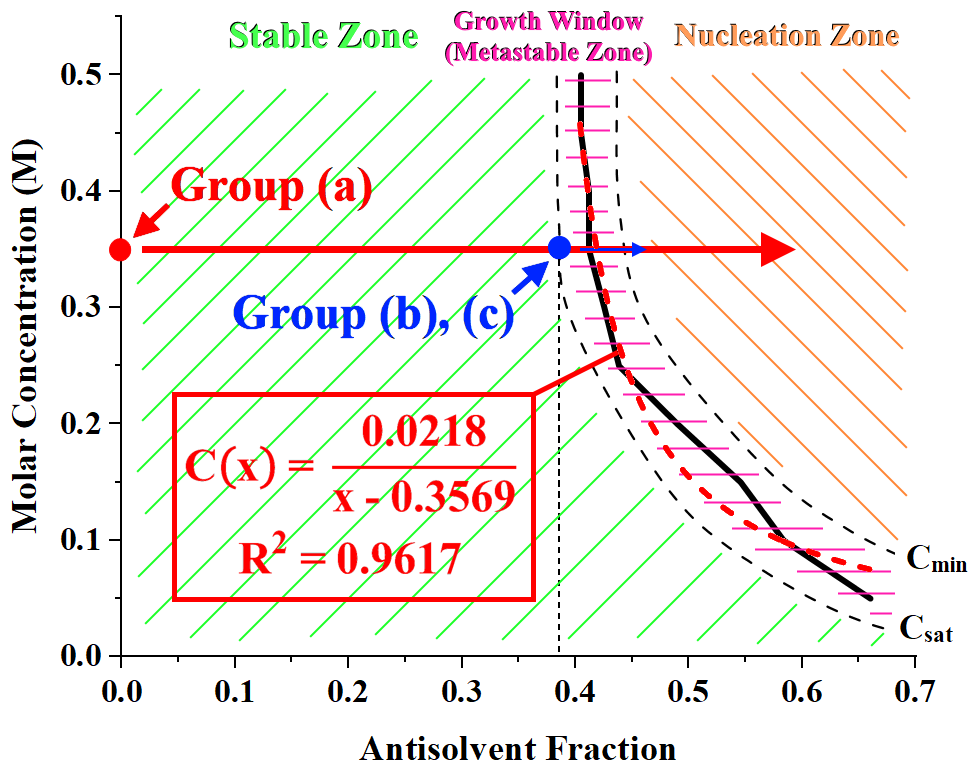}
    \caption{Solubility curve and schematic phase diagram of the CsPbBr$_3$ synthesis derived from experimental solubility data. The zoning interpretation follows the LaMer model. The "Stable Zone" (green) represents the undersaturated solution. The "Growth Window" (pink) corresponds to the metastable region ($C_{sat} < C < C_{min}$), ideal for growth of high-quality SCs. The "Nucleation Zone" (orange) indicates the region of labile supersaturation where spontaneous nucleation occurs. The red point marks the starting condition for the as-prepared solution (Group (a)), which is prone to kinetic overshoot into the nucleation zone (red arrow). The blue point marks the starting condition for the pre-treated solution (Groups (b) and (c)) after titration to $\approx$38\% ethanol fraction, positioned at the solubility boundary ($C_{sat}$), allowing for controlled entry into the metastable growth window (blue arrow). The solid black line represents the experimentally determined solubility curve, while the red dashed line shows the best fit using the empirical model presented in the inset. The equation and high R$^2$ = 0.9617 value confirm the model's accuracy.}
    \label{fgr:Fig_4}
\end{figure}

Having fixed the concentration, we next investigate the influence of the initial thermodynamic state of the solution on the nucleation and growth pathway. To interpret the experimental results, we first map the solubility limit of the precursor to define the phase diagram of the system, as illustrated in Figure~\ref{fgr:Fig_4}. The experimental solubility data, shown as a solid black line, delineate the boundary between the stable and unstable zones. These data were fitted using an empirical hyperbolic model of the form $C(x) = {A}/({x-B})$, where $C(x)$ is the molar saturation concentration and $x$ is the antisolvent fraction. The model's prediction is shown as a red dashed line, with the best-fit parameters being A = 0.0218 and B = 0.3569. The high coefficient of determination ($R^2 = 0.9617$) confirms the model's excellent agreement with the experimental measurements. This type of model is well-suited for describing solubility in binary solvent systems, where the parameter B can be physically interpreted as a critical antisolvent fraction (in this case $\approx$0.3569) below which the precursor remains highly soluble.

\begin{figure*}[!htb]
	\captionsetup{font=footnotesize}
    \centering
    \includegraphics[width=0.8\textwidth]{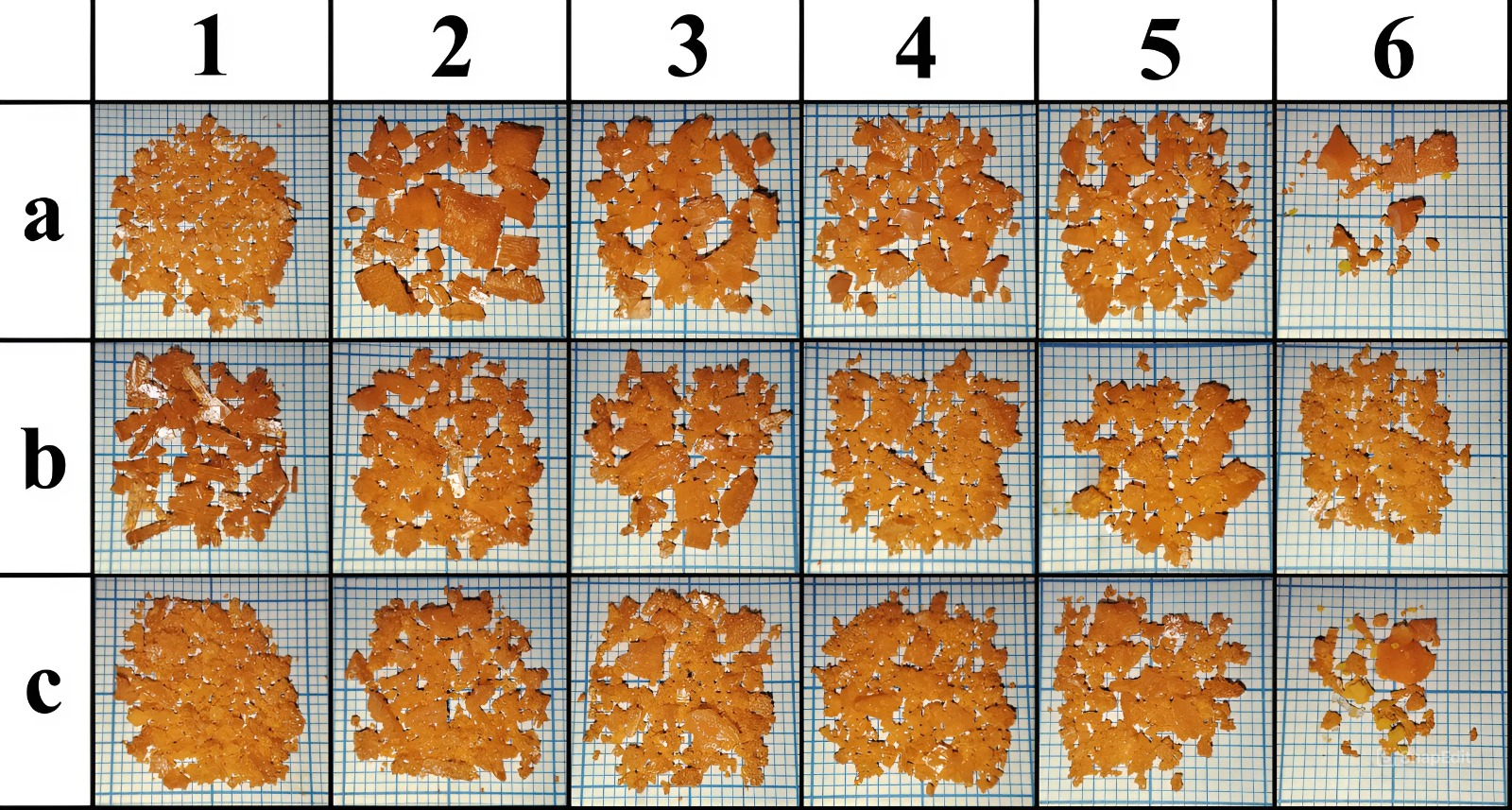}
    \caption{Photographs of CsPbBr$_3$ SCs obtained under 18 distinct experimental conditions. Rows (a-c) correspond to different precursor solution treatments: (a) as-prepared solution, (b) solution pre-treated by titration and re-filtration, (c) pre-treated solution with added seed crystals. Columns (1-6) correspond to different antisolvent compositions, with the volume percentage of DMSO in the ethanol/DMSO mixture increasing from 0\% in column 1 to 50\% in column 6, in 10\% increments.}
    \label{fgr:Fig_5}
\end{figure*}

Three distinct protocols are systematically compared based on this diagram (see growth results in Figure~\ref{fgr:Fig_5}): Group (a), using the as-prepared solution; Group (b), using a pre-treated solution; and Group (c), utilizing the pre-treated solution with seed crystals.

The results unequivocally demonstrate the superiority of the pre-treatment strategy used in Group (b). A visual comparison between Rows (a) and (b) in Column (1) of Figure~\ref{fgr:Fig_5} reveals that crystals from the pre-treated solutions are consistently larger and exhibit near-perfect crystallographic faceting. This marked improvement is directly explained by the LaMer model depicted in Figure~\ref{fgr:Fig_2}. In the standard protocol (Group (a), red point in Figure~\ref{fgr:Fig_4}), the system starts in a stable, undersaturated state. The influx of antisolvent vapor drives the concentration across the saturation threshold ($C_{\text{sat}}$) rapidly, causing it to "overshoot" the critical nucleation concentration ($C_{\text{min}}$) and triggering a Stage II burst of uncontrolled nucleation (red arrow in Figure~\ref{fgr:Fig_4}).

In contrast, the pre-titration step in Group (b) (blue point in Figure~\ref{fgr:Fig_4}) acts as a form of controlled "priming". The experimentally determined endpoint of the titration (yellow precipitate persisting for $\approx$ 10 seconds) indicates that the solution has reached the immediate vicinity of its solubility limit ($C \approx C_{\text{sat}}$). This allows the system to enter the narrow metastable growth window (pink zone in Figure~\ref{fgr:Fig_4}) smoothly, driven by the slow diffusion of antisolvent vapor (blue arrow in Figure~\ref{fgr:Fig_4}). The subsequent filtration removes any accidentally formed nuclei. Consequently, the process commences with a clean, homogeneously supersaturated solution, allowing for the orderly growth of a few nuclei without competitive precipitation.

The introduction of macroscopic (ca. \SI{1}{mm^3}) seed crystals in Group (c), which theoretically should promote growth on a designated site by lowering the energy barrier via heterogeneous nucleation~\citep{iwamatsu2011heterogeneous}, does not confer any additional advantage over Group (b) (see Figure~\ref{fgr:Fig_5}, Column (1), Row (c)). A plausible explanation is that the initial supersaturation level in the pre-treated solution is insufficient to drive growth on the seed crystal. The pre-titration followed by filtration likely brings the system to a state very close to the equilibrium saturation point ($C \approx C_{\text{sat}}$). While this condition is ideal for preventing spontaneous nucleation, it may not provide a sufficient thermodynamic driving force ($\Delta \mu > 0$) for the deposition of new layers onto the seed. In this finely balanced, near-equilibrium state, kinetic barriers to growth on the crystal facets may not be overcome, or slight temperature fluctuations could even favor dissolution over growth.

\begin{figure*}[!htb]
	\captionsetup{font=footnotesize}
    \centering
    \includegraphics[width=0.8\textwidth]{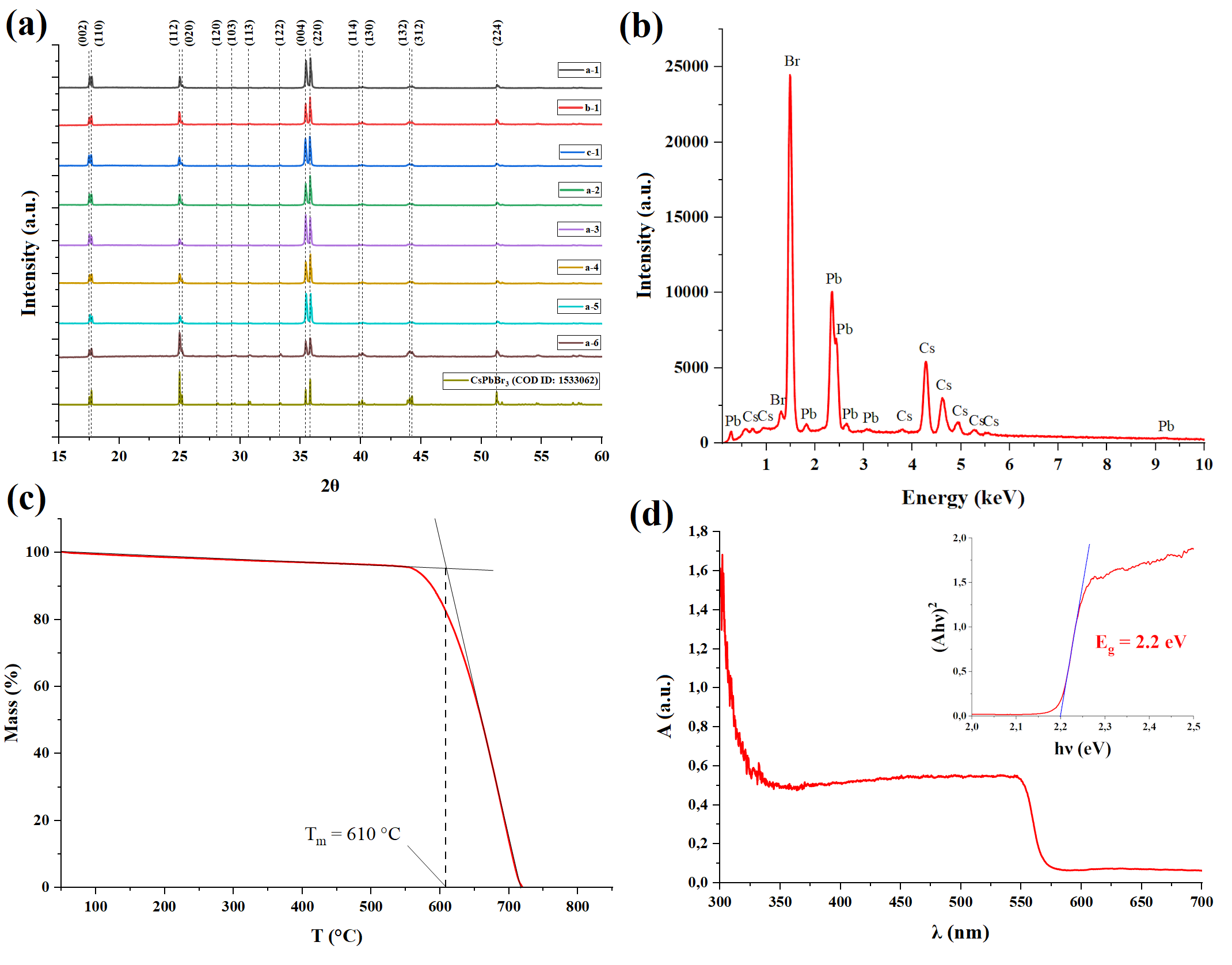}
    \caption{Comprehensive characterization of an optimized CsPbBr$_3$ SC (sample b--1). (a) Powder X-ray diffractograms of the optimized crystal (b--1) compared to representative samples from other synthesis conditions, highlighting its superior crystallinity. The reference pattern for the orthorhombic phase (COD ID: 1533062) is shown at the bottom. (b) Energy-dispersive X-ray spectrum confirming the elemental purity. (c) Thermogravimetric analysis curve showing high thermal stability. (d) UV-visible absorption spectrum and the corresponding Tauc plot (inset) for the determination of the optical band gap.}
    \label{fgr:Fig_6}
\end{figure*}

\subsubsection{Modulation of Antisolvent Diffusion Rate}

In a final optimization step, we attempt to further refine the process by slowing the kinetics of supersaturation by modulating the antisolvent diffusion rate. According to Raoult's law, adding a low-volatility component (DMSO) to the volatile antisolvent (ethanol) should decrease the partial pressure of ethanol in the vapor phase, thereby reducing its flux. Surprisingly, the experimental results are in direct opposition to this expectation (see Figure~\ref{fgr:Fig_5}, Column (2)--(6)). Across all three groups, increasing the DMSO fraction in the antisolvent reservoir leads to a consistent and dramatic reduction in both crystal yield and quality. This counterintuitive outcome strongly suggests the presence of an inhibiting secondary mechanism. We hypothesize that this is due to the formation of a diffusion barrier at the solution-gas interface. Although DMSO has a low vapor pressure, a small amount could co-evaporate and subsequently condense on the surface of the precursor solution. Since DMSO is an excellent solvent for CsPbBr$_3$, this process would create a thin, stable, solvent-rich film at the interface. This film would then act as a physical barrier, effectively shielding the bulk precursor solution from the incoming ethanol antisolvent vapor. The inefficient and non-uniform mass transport would cripple the process of supersaturation, suppressing both nucleation and growth.

In summary, this systematic experimental investigation, which is designed based on the principles of crystallization theory, allows us to identify an optimal set of parameters. The protocol corresponding to sample \textbf{b--1} -- a \SI{0.35}{M} precursor solution, 9:1 (v/v) DMSO/DMF solvent mixture, pre-treatment by ethanol titration to a metastable state, and pure ethanol as the antisolvent -- provides the most reliable and effective pathway for the reproducible synthesis of large, high-quality CsPbBr$_3$ SCs.

\subsection{Characterization of Optimized CsPbBr$_3$ Single Crystals}

To validate the efficacy of our theory-guided synthesis protocol, the CsPbBr$_3$ SCs obtained under the optimized conditions (sample b--1) are subjected to a comprehensive structural, compositional, thermal, and optical analysis.

\subsubsection{XRD analysis}
The crystal structure and phase purity are assessed by powder X-ray diffraction (XRD). It is confirmed that all synthesis protocols yield phase-pure orthorhombic CsPbBr$_3$ (space group \textit{Pbnm}, COD ID 1533062), regardless of the growth kinetics (Figure~\ref{fgr:Fig_6}a). Since the analysis is performed on pulverized samples, these results validate the chemical composition but do not reflect macroscopic parameters such as crystal size or optical transparency. Therefore, while the correct phase formation is achieved in all regimes, the superiority of the optimized protocol (b--1) lies in the macroscopic quality of the resulting crystals rather than the phase purity alone. To complement the bulk powder analysis and assess the structural quality of the individual SCs obtained by optimized protocol (b--1), SC-XRD measurements were subsequently performed. With the chosen exposure time of 20 seconds per frame, 126 reflections were collected during the measurement, of which 97 (77\%) were integrated into an orthorhombic unit cell with parameters a = \SI{8.241(5)}{\angstrom}, b = \SI{8.226(16)}{\angstrom}, c = \SI{11.79(3)}{\angstrom}, V = \SI{799(3)}{\angstrom}$^{3}$, confirming the low-temperature orthorhombic modification of CsPbBr$_{3}$~\citep{rodova2003large, xian2020emerging, balvanz2024unveiling}.

\begin{table*}
  	\caption{Quantitative EDX analysis of the CsPbBr$_3$ SC based on the Cs-L$\alpha$, Pb-L$\alpha$, and Br-K$\alpha$ lines. Calculated average values are based on five measurements.}
  	\footnotesize
  	\setlength{\tabcolsep}{4pt}
  	\label{tbl:Table_2}
  	\begin{tabular}{@{}ccccc@{}}
    \toprule
    \textbf{No.} & \textbf{Cs (at.\%)} & \textbf{Pb (at.\%)} & \textbf{Br (at.\%)} & \textbf{Ratio (Cs : Pb : Br)} \\
    \midrule
    1 & 20.87 & 19.91 & 59.21 & 1.05 : 1.00 : 2.97 \\
    2 & 20.69 & 19.19 & 60.12 & 1.08 : 1.00 : 3.13 \\
    3 & 20.77 & 19.54 & 59.69 & 1.06 : 1.00 : 3.05 \\
    4 & 20.27 & 19.04 & 60.69 & 1.06 : 1.00 : 3.19 \\
    5 & 20.38 & 18.93 & 60.69 & 1.08 : 1.00 : 3.21 \\
    \midrule
    \multicolumn{1}{l}{\textbf{Average}} & \textbf{20.60 $\pm$ 0.26} & \textbf{19.32 $\pm$ 0.40} & \textbf{60.08 $\pm$ 0.65} & \textbf{1.07 : 1.00 : 3.11} \\
    \bottomrule
  \end{tabular}
\end{table*}

\subsubsection{EDX analysis}
The elemental composition of the optimized crystal is determined by EDX. Given the unpolished nature of the as-grown SCs, the choice of analytical lines is critical. As revealed by SEM imaging (see Figure~\ref{fgr:Fig_S1} in Supplementary Information), the crystal surface exhibits significant topography, including growth steps and terraces. Standard analysis using low-energy X-ray lines (Br L$\alpha$ and Pb M$\alpha$, see Figure~\ref{fgr:Fig_6}b) is highly susceptible to geometric shadowing and self-absorption by these surface irregularities, leading to significant quantification errors. Therefore, to ensure accuracy, quantitative analysis is performed exclusively using high-energy lines (Br K$\alpha$ and Pb L$\alpha$), which possess greater penetrating power and are insensitive to surface morphology.

The results obtained using this optimized methodology are summarized in Table~\ref{tbl:Table_2}. The analysis yields an atomic ratio of Cs:Pb:Br $\approx$ 1.07:1.00:3.11. This composition is remarkably close to the ideal 1:1:3 stoichiometry. The slight remaining excess of Cs and Br requires explanation. One direct cause is the presence of trace secondary phase micro-inclusions confined to the surface. As shown in Figure~\ref{fgr:Fig_S2} (see Supplementary Information), while the crystals appear visually homogeneous and display the characteristic orange color of the perovskite phase under ambient light, localized green luminescent spots are visible only under UV illumination ($\lambda = 365$ nm). These spots are attributed to the Cs$_4$PbBr$_6$ phase, which is known to form under similar synthesis conditions~\citep{chen2016large}. Furthermore, as our synthesis operates at the boundary of a metastable growth window ($K=1.5$), it is possible that the growth conditions promote the formation of intrinsic point defects within the lattice. Theoretical studies suggest that defects such as cesium interstitials (Cs$_i$) or lead vacancies ($V_{Pb}$) can become thermodynamically favorable under certain conditions, which would also contribute to the observed minor non-stoichiometry~\citep{kang2017high}.

\subsubsection{TGA analysis}
Thermal stability is a critical parameter for the long-term operational reliability of optoelectronic devices. TGA is performed to evaluate this property, with the results shown in Figure~\ref{fgr:Fig_6}(c). The TGA curve demonstrates the excellent thermal stability of the synthesized SC. No significant mass loss is observed from room temperature up to approximately \SI{550}{\celsius}, which indicates the absence of trapped residual solvent molecules within the crystal lattice and the intrinsic structural integrity of the material. Above this temperature, a sharp, single-step decomposition process begins, with an onset temperature of rapid mass loss at T$_m$ = \SI{610}{\celsius}. This high decomposition temperature confirms that the crystals are suitable for applications requiring high thermal and operational stability.

\subsubsection{UV-VIS analysis}
Finally, the optical quality of the optimized crystal is investigated using UV-visible absorption spectroscopy. The spectrum, presented in Figure~\ref{fgr:Fig_6}(d), is characterized by a very sharp absorption edge located at approximately \SI{560}{\nano\meter} and a flat, near-zero absorption baseline in the sub-bandgap region. This steep onset is a hallmark of a direct band gap semiconductor with a low density of defect-induced states (Urbach tail), indicative of high material quality. To precisely determine the optical band gap ($E_g$), the Tauc relation for a direct allowed transition is applied~\citep{tauc2012amorphous}:
\begin{equation}
    (\alpha h\nu)^2 = A(h\nu - E_g)
    \label{eqn:Eq_7}
\end{equation}
where $\alpha$ is the absorption coefficient, $h\nu$ is the photon energy, and $A$ is a constant. The Tauc plot of $(\alpha h\nu)^2$ versus $h\nu$ is shown in the inset of Figure~\ref{fgr:Fig_6}(d). By extrapolating the linear portion of the curve to the energy axis intercept, the optical band gap is determined to be $E_g = \SI{2.20}{\electronvolt}$. This value is in excellent agreement with the band gaps reported for high-quality CsPbBr$_3$ SCs in the literature~\citep{fan2020solution, zhang2019high, su2024growth, ding2017high, rakita2016low}, further validating the high optical and electronic quality of the material synthesized via our rational design approach.


\section{Conclusion}

In this work, we have presented a systematic methodology for the synthesis of large, high-quality CsPbBr$_3$ SCs, where experimental design is consistently informed by theoretical principles. Moving beyond purely empirical approaches, our study demonstrates a synergy between fundamental theory and targeted experimentation.

The process began by establishing a rational chemical space for experimentation. A synergistic 9:1 (v/v) DMSO/DMF binary solvent was selected to balance the requirements of high precursor solubility and controllable crystallization kinetics, a choice rationalized by an analysis of Gutmann's Donor Numbers. Subsequently, ethanol was selected from a pool of candidates by evaluating its properties against key theoretical criteria of miscibility and controlled mass transport rate using Hansen Solubility Parameters and vapor pressure.

Within this rationally-defined chemical system, a systematic experimental optimization, guided by the qualitative principles of classical nucleation and growth theories, was performed to map the optimal "growth window". This involved identifying a precursor concentration of \SI{0.35}{M} as an ideal compromise and implementing a crucial preliminary titration step to guide the system into a productive metastable state, thereby suppressing uncontrolled homogeneous nucleation.

The resulting optimized protocol consistently yields phase-pure, orthorhombic CsPbBr$_3$ SCs up to \SI{1}{cm} in size within one week at room temperature. Comprehensive characterization confirmed the high quality of these crystals, which exhibit excellent crystallinity, robust thermal stability up to \SI{550}{\celsius}, a sharp absorption edge located at approximately \SI{560}{\nano\meter}, and an optical band gap of \SI{2.20}{\electronvolt}. This work provides a case study in how fundamental principles of solution chemistry and crystallization can be applied to move beyond purely empirical optimization, offering a more systematic approach to protocol design for producing high-quality perovskite SCs, paving the way for their broader utilization in stable and efficient optoelectronic devices.

\section*{Acknowledgement}

We thank A. N. Nekrasov, G. V. Kiryukhina, D. A. Chareev and A. B. Yeresko for the data obtained by means of the scanning electron microscopy, single-crystal X-ray diffraction and thermogravimetric analysis. The research was carried out within the state assignment of the Russian Science Foundation (project No. 23-19-00884).

\section*{Author contributions}
I.O. S.: Conceptualization, Investigation, Data Curation, Visualization, Writing – Original Draft; R.G. N.: Methodology, Conceptualization, Supervision, Validation, Writing – Review \& Editing; T.N. V.: Formal analysis, Investigation, Writing – Review \& Editing.

\bibliographystyle{apsrev4-2} 
\bibliography{references}

\clearpage
\appendix
\onecolumngrid
\setcounter{figure}{0}
\renewcommand{\thefigure}{S\arabic{figure}}

\section*{Supplementary Information}

\begin{figure}[h!]
	\captionsetup{font=footnotesize}
    \centering
    \includegraphics[width=0.8\textwidth]{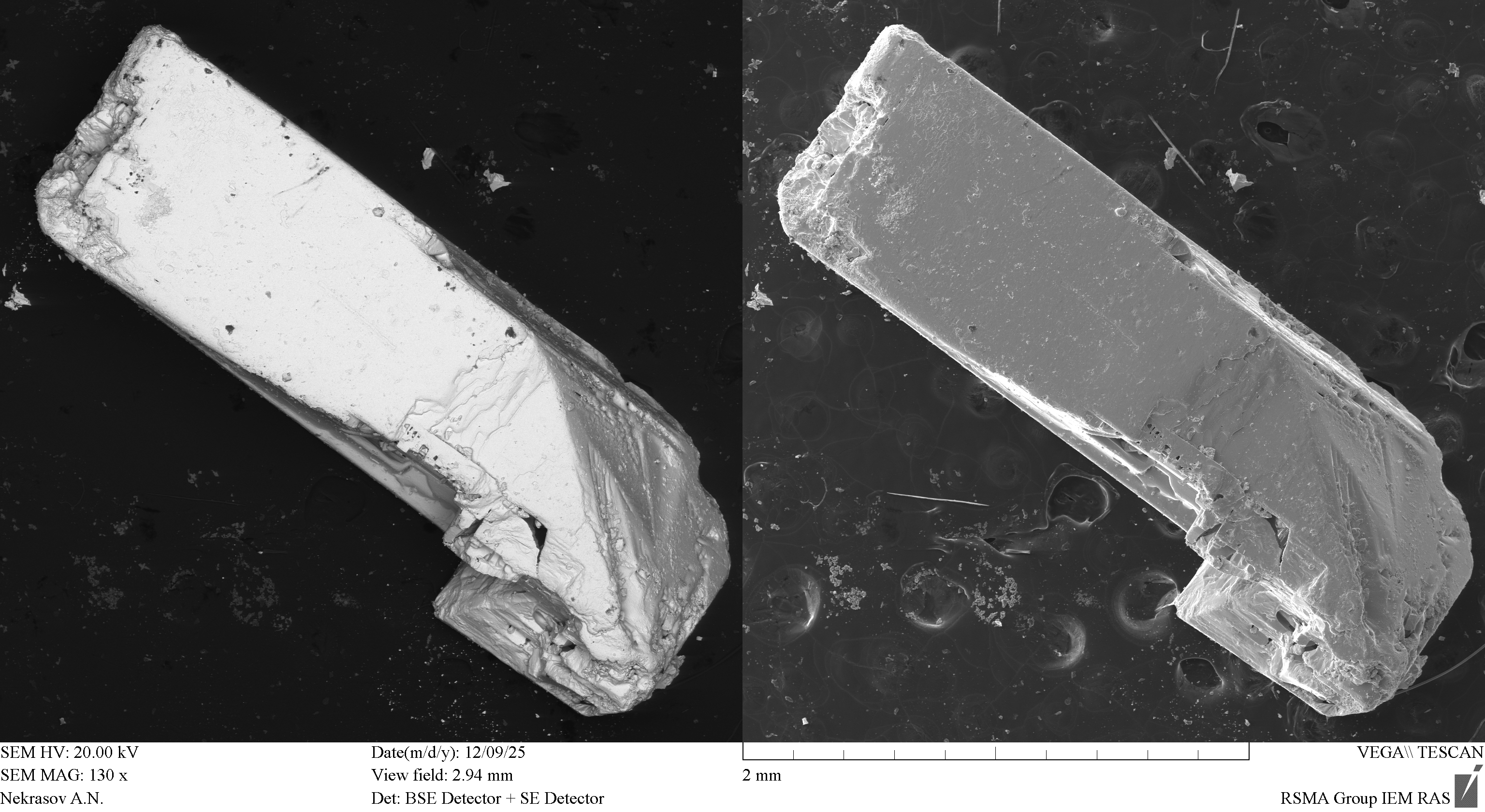}
    \caption{Scanning Electron Microscopy (SEM) images of a representative unpolished CsPbBr$_3$ single crystal used for EDX analysis. The left panel displays the Backscattered Electron (BSE) image, which exhibits uniform contrast across the crystal, indicating a high degree of compositional homogeneity without segregation of heavy or light elements. The right panel shows the Secondary Electron (SE) image of the same region, revealing significant surface topography, including growth steps and terraces. This observed surface roughness confirms the necessity of utilizing high-energy X-ray lines (Br K$\alpha$, Pb L$\alpha$) for accurate quantitative analysis to avoid geometric shadowing effects associated with soft radiation.}
    \label{fgr:Fig_S1}
\end{figure}

\vspace{0.5cm}

\begin{figure}[h!]
	\captionsetup{font=footnotesize}
    \centering
    \includegraphics[width=0.8\textwidth]{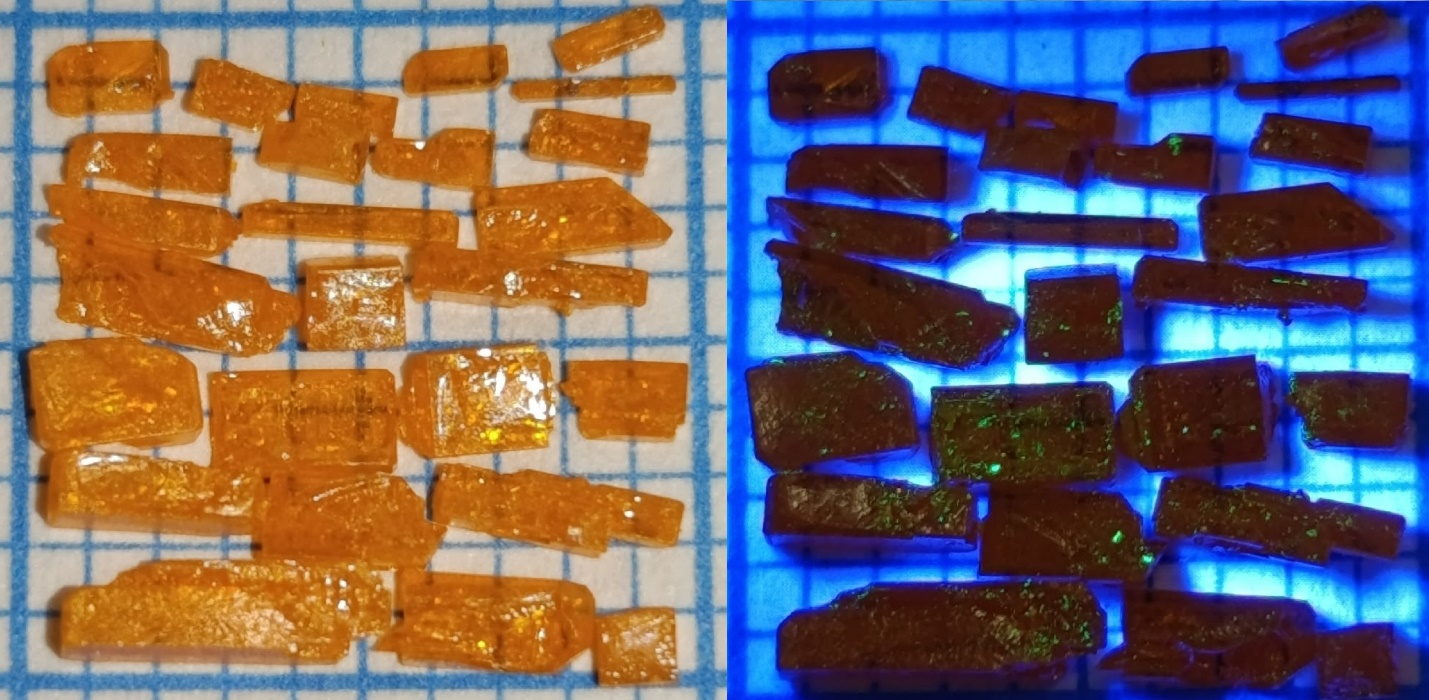}
    \caption{Optical photographs of the grown CsPbBr$_3$ crystals placed on millimeter grid paper. The left panel shows the crystals under ambient daylight, displaying the characteristic orange color and visual homogeneity of the perovskite phase. The right panel shows the same crystals under UV illumination ($\lambda_{ex} = 365$ nm). While the bulk of the crystals remains non-emissive, distinct localized green luminescence spots are observed on the surfaces. This specific green emission is attributed to the formation of trace amounts of the Cs$_4$PbBr$_6$ zero-dimensional (0D) phase, confirming that secondary phase inclusions are confined to the surface rather than the bulk.}
    \label{fgr:Fig_S2}
\end{figure}

\end{document}